\documentclass{article}

\usepackage{microtype}
\usepackage{graphicx}
\usepackage{subfigure}
\usepackage{booktabs} 

\usepackage[hyphens]{url}
\usepackage{hyperref}

\usepackage[accepted]{icml2020}

\icmltitlerunning{AligningRecommenderSystems}

\begin{document}

\twocolumn[
\icmltitle{What are you optimizing for?\\  Aligning Recommender Systems with Human Values}



\icmlsetsymbol{equal}{*}

\begin{icmlauthorlist}
\icmlauthor{Jonathan Stray}{pai}
\icmlauthor{Ivan Vendrov}{cons}
\icmlauthor{Jeremy Nixon}{cons}
\icmlauthor{Steven Adler}{pai}
\icmlauthor{Dylan Hadfield-Menell}{ed,chai}
\end{icmlauthorlist}

\icmlaffiliation{pai}{Partnership on Artificial Intelligence, San Francisco, California, USA}
\icmlaffiliation{ed}{Department of Electrical Engineering and Computer Science, UC Berkeley, California, USA}
\icmlaffiliation{chai}{Center for Human-Compatible Artificial Intelligence, Berkeley, California, USA}
\icmlaffiliation{cons}{Author names were withheld in previously published versions of this paper by determination of their employer.}

\icmlcorrespondingauthor{Dylan Hadfield-Menell}{dhm@eecs.berkeley.edu}

\icmlkeywords{Machine Learning, ICML}

\vskip 0.3in
]

\printAffiliationsAndNotice{} 

\begin{abstract}
We  describe cases where real recommender systems were modified in the service of various human values such as diversity, fairness, well-being, time well spent, and factual accuracy. From this we identify the current practice of values engineering: the creation of classifiers from human-created data with value-based labels. This has worked in practice for a variety of issues, but problems are addressed one at a time, and users and other stakeholders have seldom been involved. Instead, we look to AI alignment work for approaches that could learn complex values directly from stakeholders, and identify four major directions:   useful measures of alignment, participatory design and operation, interactive value learning, and informed deliberative judgments.

\end{abstract}

\section{Introduction}
Recommender systems are AI systems that present users with a tailored set of items based on factors such as past user behavior, user attributes, and features of the underlying items. They rank huge numbers of items in order to determine which content each user sees or interacts with on social media feeds, video platforms, news aggregators, and in online stores.



Classically, recommender designers have framed their goal as providing “relevant” items to each user. As these technologies have rolled out into society, it has become clear that the appropriate goals are often highly context-dependent, nuanced, and consequential. Previous work has argued for the incorporation of additional objectives including the promotion of accurate information \citep{Stocker2019}, fairness \citep{barocas2017fairness}, and diversity of content \citep{castells2015novelty}, and the reduction of addiction \citep{hasan2018excessive}\cite{Andreassen2015} and polarisation \citep{benkler2018network}. These represent values-driven decisions about what `we,' the relevant stakeholders, do and don't want recommended and how items  should be ranked. 

There are further, less-explored opportunities for values-driven design. A specification for a recommender system could include, e.g., that a news recommendation system to support values like informedness, autonomy, inclusiveness, participation, decentralization, representation, deliberation, and tolerance \citep{helberger2019} or that social media recommenders “provide positive incentives for encouraging more constructive conversation” \citep{Wagner2019}. Recommenders for books, music and video might want to promote cultural exploration and ensure that emerging artists gain exposure \citep{Mehrotra2018}. An online store may want its product recommendations to encourage the sale of products with a smaller carbon footprint \citep{rolnick2019tackling} or otherwise support social and environmental goals. Recommenders have already been built around high-level goals like improving user well-being \citep{khwaja2019aligning}\citep{Stray2020} and even proposed as an aid to self-actualization \citep{Knijnenburg2016}.


In this paper, we propose that recommender system design can draw from the study of \emph{the value alignment problem}: the problem of ensuring that an AI system's behavior aligns with the values of the \emph{principal} on whose behalf it acts \cite{hadfieldmenell2019incomplete}. In the corresponding \emph{recommender alignment problem}, the objective is to align recommendations with the goals of users, system designers, and society as a whole. In Section 2, we ground the recommender alignment problem with case studies of how the builders of large recommendation systems have responded to domain-specific challenges. We observe a common three phase approach to alignment: 1) relevant categories of content (e.g., clickbait) are \emph{identified}; 2) these categories are \emph{operationalized} as evolving labeled datasets; and 3) models trained off this data are used to adjust the system recommendations. In Section 3, we draw from AI alignment work to propose high-level approaches to the design of recommendation systems whose output embodies human values in a more nuanced, participatory and adaptable way.  We do not attempt to catalog the ethical issues surrounding recommenders (see e.g. \cite{Milano2019}) or to say how various values should be prioritized. Rather, we are interested in developing \emph{affordances} that allow high-level, inclusive control of what a recommendation system recommends.



\section{Case Studies}

This section aims to ground our discussion of higher-level approaches to recommender alignment in concrete examples of diverse ways that large commercial recommender systems have been modified to be “better."

\subsection{Identifying unwanted content}
Clickbait is one of the earliest recognized examples of recommender misalignment. Early recommender systems were often optimized to show the user the items they would be most likely to click on, which rewarded items with attention-grabbing titles but not necessarily quality content. In response, system designers identified alternate metrics that are harder to game and incorporated them into system training. For example, “dwell time”, the amount of time a user spends on a page before coming back to recommender results, has been used as ranking and personalization signal at many companies including Yahoo \cite{Yi2014} and Facebook\cite{KhalidEl-Arini2014}. Other simple metrics, such at the click-to-share ratio on Facebook posts \cite{KhalidEl-Arini2014}, have also been used in practice to suppress clickbait. 

However, hand-designed metrics are typically unable to capture the complexity of even relatively simple and uncontroversial concepts. Most large production systems today (including Flipboard \cite{Cora2017} and Facebook \cite{Peysakhovich2016}) instead train classifiers on human-labelled clickbait and use them to measure and decrease the prevalence of clickbait within the system. Recommenders that use these learned classifiers are more effective than those that use only hand-designed metrics. As we discuss below, we attribute this to the combination of a greater quantity of human feedback with a more flexible model.

There are a number of other types of content that one might not want to recommend broadly, such as hateful or harassing posts, sexual content, posts that advocate violence, or misinformation. These are too complex to detect with simple metrics derived from user behavior. Their precise definition is politically contested, and the decision to remove or downrank this content must balance potential harm, local laws, and freedom of expression in complex, context-sensitive ways \cite{York2019}. Hateful or harassing posts are challenging to identify by machine because of the nuances of language and context but sometimes significant  is available, for example if a comment occurs within a larger online discussion. A “toxicity” model created by Jigsaw, trained on human-labelled examples, has been tested at Twitter as part of a “healthy conversations” metric \cite{Wagner2019}, and used by the New York Times for comment triage prior to human review \cite{Etim2017}. However, such classifiers can misunderstand the significance of utterances in race- and gender-biased ways \cite{Salganik2020}.

Mis- and disinformation is a much harder problem for automated detection because it may require open-ended research, e.g. verifying a video by identifying and speaking to a witness \cite{Silverman2020}. Nonetheless, major platforms including Facebook \cite{TessaLyons2018} now employ trained models in various aspects of their counter-misinformation efforts. These typically operate at the level of classifying domains, users, or near-duplicates of previously debunked items \cite{Cora2017} \cite{Facebook2020}. The final source of truthfulness is human ratings of individual articles, photos, and videos. At Facebook this information is provided both by a network of professional fact-checking organizations and by users who flag content as “false news” \cite{Thompson2018}. 

\subsection{Promoting desirable outcomes}
More recently, recommender system builders have attempted to identify content that furthers particular values-based goals. Goals addressed in practice include healthy public conversations, exposure for little-known creators or sellers, and the well-being of social media users. Rather than just identifying more or less desirable items, these richer conceptions of recommender alignment envision the support of positive values. Recommender systems could support democratic values in a variety of ways \cite{helberger2019}, and encourage well-being \cite{khwaja2019aligning}, personal self-actualization \cite{Knijnenburg2016}, and human flourishing \cite{Rodriguez2009}.

Although the empirical evidence for recommender systems' role in creating “filter bubbles” is weak \cite{Bruns2019},  most discussions of recommenders and the public sphere hold diversity of viewpoint as an essential value \cite{Helberger2018}. Such studies typically consider diversity with respect to marginalized communities or left-right politics, but there are many other kinds of content diversity that recommender systems could promote such as topic or cultural diversity, and various concepts of representative fairness \cite{Gao2020}.  


Spotify is notable for elaborating on the fairness and diversity issues faced by a music platform. Their recommendation system is essentially a two-sided market, since artists must be matched with users in a way that satisfies both, or neither will stay on the platform. Recommender systems (and human cultural markets in general) often suffer from “superstar economics” where a small number of stars control the majority of the attention \cite{Mehrotra2018}. The company has also discussed the representation of different demographics, and the value of musical exploration, leading to a domain-specific definition: a playlist is “fair” if it contains songs from artists at different levels of popularity \cite{Lalmas-Roelleke2019}. This is a concrete example of the management of a multi-stakeholder recommendation system, a central topic in recommender alignment which is starting to be more explored \cite{Abdollahpouri2019}.

Especially when filtering social media and news content, recommenders are key mediators in public discussion. Twitter has attempted to create “healthy conversation” metrics with the goal to “provide positive incentives for encouraging more constructive conversation” \cite{Wagner2019}. One conception of these metrics includes shared attention, shared reality, variety and receptivity, and a prototype has been built with models trained on human labels \cite{Gadde2018}. The Dutch FG Mediagroep modified their news recommender to increase the consumption of "timely and fresh" content, as well as overall coverage, defined as the proportion of articles that were recommended to any user\cite{Lu2020}

In the context of a mobile health app, researchers at Telefonica have shown that personalizing recommendations based on the user’s personality type, e.g. recommending social activities to extroverts, could result in a meaningful improvement in well-being outcomes \cite{khwaja2019aligning}.  Facebook modified several of its recommendation systems in 2017-2018 to include consideration of "meaningful social interactions," an explicit proxy for well-being \cite{Stray2020}. Such an intervention has potentially large effects because there are so many Facebook users, but no outcome information has so far been published.

\subsection{Characterizing the State-of-the-Art}
These case studies illustrate a common pattern in current recommender efforts: identifying and operationalizing abstract concepts like “healthy conversation” or “clickbait”, then adjusting the recommendation algorithm to change the observed prevalence of these concepts.


\textbf{Identification:} This is the phase where system designers become aware of a negative outcome associated with the system and identify a concept associated with it. For example, identifying that users are getting drawn into misrepresented videos for short periods of time and developing a definition of `clickbait’.

\textbf{Operationalization:} A concrete procedure is developed to identify instances of the abstract concept in the recommender system. Although this can be done through manual review, most systems rely on some form of machine learning in order to scale. These models are typically trained from an evolving dataset where labels are provided by human workers or observable user behavior (e.g., flagging spam, responses to user surveys).

\textbf{Adjustment:} In this step, system designers modify the recommender system in order to increase or decrease the prevalence of the desired concept, as measured by the classification process defined in step 2. This can be done in a number of ways, most commonly: 1) \textit{At training:} Adding concept prevalence as an additional training objective in a multi-objective training setup, as in \cite{Zhao2019}; 2) \textit{At launch:} Tracking concept prevalence as a top-line metric and preferentially launching changes that improve it; 3) \textit{At serving:} Directly add or remove items from the recommendation pool or change their rankings based on their classification, possibly after manual review.

\section{Higher-level approaches to recommender alignment}
A major problem with current practice is that it is fundamentally reactive. Problems are addressed one at a time, often after they are widespread. This is most troublesome when addressing one problem causes a new and harder-to-quantify problem, e.g. moving from optimizing for clicks to optimizing for time-in-app may reduce clickbait but could increase addiction. Even when there’s no risk of creating new problems, this approach simply cannot scale to handle the interests of billions of people embedded in millions of communities. Conducting detailed user research, designing new metrics, creating appropriate data sets, and training classification models is an iterative process that takes months if not years. 

Moreover, the bottleneck in the problem identification step implies that only the problems most visible to system designers get addressed. This privileges issues that affect system designers and their social circles. Even more troubling is that it privileges the system designers' moral and ethical views. We wish to address the problem of encoding appropriate values into recommender systems in a more general, participatory, and scalable way.

\subsection{Useful definitions and measures of alignment}
There is no single definition of value that a recommendation system should adhere to. This is in part a philosophical problem, in part the reality of political pluralism, and in part because value depends tremendously on context~\cite{gabriel2020artificial}. But there are still many values that are widely shared, and useful instruments to evaluate human outcomes.

Suites of metrics for “well-being” are used in public policy ~\cite{ODonnell2014} and many of those most applicable to AI systems have been collected into an IEEE standard for well-being assessment~\cite{schiff2020ieee}. Metrics for other high-level values, such as agency, have been developed in fields such as development economics ~\cite{Alkire2008}. These assessments rely on a combination of self-reported subjective states (“I am tired”), self-reported objective states (“I went to work today”), and statistics (“The unemployment rate is 5.4\%”) ~\cite{OECD2018}. 

Many other metrics have been developed to evaluate recommenders, but that leaves open the question of which measures are most useful for value alignment in each domain. Standard evaluation data sets or protocols for recommenders in a particular domain will greatly accelerate progress on recommender alignment.

 \subsection{Participatory recommenders}
The operators of recommender systems have often worked in a reactive mode, responding to issues after they are identified, and in the way that makes the most sense to them. It would be better if users could be effectively involved in the initial design of the system, had more control over routine operation, and good ways to escalate new issues.

Most real recommender systems must trade off between users or consumers, producers or sellers, and the recommender operator~\cite{Abdollahpouri2019}. Recommendations may also affect non-users in a variety of ways, such as product recommendations that have adverse environmental consequences or driving directions that create congestion in otherwise calm neighborhoods~\cite{kulynych2020pots}. There sorts of issues are related to the classic economic problems of externalities and public goods, and are explicitly considered in multi-stakeholder recommendation research~\cite{milano2019ethical}. Users also have preferences around what recommendations \emph{other} users see, such as when a someone flags something as misinformation, or when people work together to up- and down-vote items ~\cite{York2019}.

The challenge here is in figuring out who the stakeholders are, how to involve them, and how to resolve their competing demands ~\cite{Baum2020}. One participatory design effort created an online matching system for a volunteer food delivery service, attempting "to enable stakeholders to construct a computational model that represents their views and to have those models vote on their behalf to create algorithmic policy"~\cite{Lee2019}. Volunteer drivers, donors, recipients, and non-profit employees collaboratively designed the trade-offs between themselves and between efficiency and equity. This was done by interactive elicitation and modeling of individual preferences, followed by a Borda voting rule to decide the choice for the overall system. The authors report that this new algorithm both improved distributive outcomes and was perceived as fairer.

\subsection{Interactive value learning}
When users make more of the decisions, or when their intent is more clearly expressed, such as in search engines or e-commerce sites, many of the problems with recommenders are alleviated. 
The recommenders with the most troubling side effects often have passive, low-agency, low-bandwidth interfaces like infinitely scrolling content feeds or autoplay. In these systems, user traces are evaluated by a combination of hand-coded and learned metrics. It is often unclear, even to system designers, what 'values' these metrics represent, as optimization-driven behavior can be counter-intuitive~\cite{krakovna_2018}.

Direct user control of how values are enacted would provide both agency and transparency. This implies that recommenders must learn values interactively. There is a general sentiment that offline recommender benchmarks are decreasingly meaningful ~\cite{Rohde2018}, which further suggests that static values training data sets are inadequate. 

Rather than depending on surveys for user feedback, we could be thinking about new kinds of ongoing interaction protocols. For example, probabilistic models like Inverse Reward Design~\cite{hadfieldmenell2017inverse} can be used to regularize recommender metrics and infer consistent combinations of metrics that generalize intent~\cite{ratner2018simplifying}. Interfaces designed around efficient imitation learning~\cite{brown2020safe} and active reward learning~\cite{biyik2020active} might be a path towards more accessible, nuanced and predictable control of recommendation behavior.

\subsection{Design around informed, deliberative judgment}



Converging arguments in AI alignment ~\cite{Tarleton2010}, political philosophy ~\cite{sep-reflective-equilibrium}, and bioethics ~\cite{Molewijk2008} suggest that informed, deliberative, and perhaps retrospective evaluations are of a higher quality than immediate judgements. Many people report that they watch more TV than they retrospectively endorse ~\cite{Frey2007} and a similar type of behavior happens with some social media users ~\cite{Andreassen2015}. The notion of "time well spent" gets at similar ideas.

This is closely related to the common problem of focusing on short vs. long term outcomes, which is an issue with the use of metrics in AI generally ~\cite{Thomas2020}. One solution is to solicit informed, retrospective feedback. For example, one could show the user a summary of their recommendations and usage patterns in the past month, ask them to spend some time reflecting on it, and then rate how happy they are with this or how well it matches their goals. Versions of this idea have seen some use in the tech industry, sometimes known as "the regret test" ~\cite{Eyal2018}.

\section{Conclusion}
Recommendations systems are a cornerstone of the internet economy. They allow people to effectively interact with the gigantic amount of information found in modern internet platforms. As recommendation systems proliferate, it is important to consider the impact these systems have and the values that drive ranking decisions. A crucial step in this process is to understand how these values are currently designed and updated. As our case studies illustrate, this can be largely described with a three step process where 1) outcomes are identified and reified as concepts; 2) the concepts are operationalized with a hand-coded or learned metric; and 3) this metric is used to adjust recommendation behavior. Looking to the future, we are excited about the potential for recommendation systems to better align with human values through the incorporation of well-being metrics, participatory approaches to objective design, interactive value learning, and optimizing for informed and deliberative preferences. 

\bibliography{aligning_recommenders}
\bibliographystyle{icml2020}

\end{document}